\documentclass[%
aps,
jmp,%
amsmath,amssymb,
reprint,%
]{revtex4-1}
\usepackage{graphicx}
\usepackage{multirow}
\usepackage{diagbox}
\usepackage{subfigure}
\usepackage{dcolumn}
\usepackage{bm}
\usepackage[mathlines]{lineno}
\usepackage{epstopdf}
\usepackage{amsmath}
\usepackage[colorlinks]{hyperref}
\hypersetup{
	colorlinks = true,
	citecolor = {blue},
	linkcolor = {red}
}

\newcommand{\blk}{\color{black}}
\newcommand{\red}{\color{red}}
\begin{document}
	\title{Experimental secure quantum key distribution in presence of polarization-dependent loss}
	\author{Chunfeng Huang,$^{1}$ Ye Chen,$^{1}$ Long Jin,$^{1}$ Minming Geng,$^{2}$ Junwei Wang,$^{3}$ Zhenrong Zhang,$^{2}$ and Kejin Wei$^{1,*}$ }
	
	\address{
		$^1$Guangxi Key Laboratory for Relativistic Astrophysics, School of Physical Science and Technology,
		Guangxi University, Nanning 530004, China\\		$^2$Guangxi Key Laboratory of Multimedia Communications and Network Technology, School of Computer, Electronics, and Information, Guangxi University, Nanning 530004, China\\
		$^3$CAS Quantum Network Co., Ltd, Shanghai 201315, China\\
		$^*$Corresponding author: kjwei@gxu.edu.cn
	}
	\date{\today}
	
	\begin{abstract}
		
		Quantum key distribution (QKD) is theoretically secure using the principle of quantum mechanics; therefore, QKD is a promising solution for the future of secure communication. Although several experimental demonstrations of QKD have been reported, they have not considered the polarization-dependent loss in state preparation in the key-rate estimation. In this study, we experimentally characterized polarization-dependent loss in realistic state-preparation devices and verified that a considerable PDL exists in fiber- and silicon-based polarization modulators. Hence, the security of such QKD systems is compromised because of the secure key rate overestimation. Furthermore, we report a decoy-state BB84 QKD experiment considering polarization-dependent loss. Finally, we achieved rigorous finite-key security bound over up to 75 km fiber links by applying a recently proposed security proof. This study considers more realistic source flaws than most previous experiments; thus, it is crucial toward a secure QKD with imperfect practical devices.

	\end{abstract}
	
	\maketitle
	
	\section{introduction}
	Quantum key distribution (QKD) has received great interest as it is an information-theoretic security communication technology~\cite{1999Lo}. With much effort, QKD has been experimentally demonstrated over  fiber-based~\cite{2016Choi,2017Canas,2018Boaron2,Liuyang2010,2012wangshuang,2021Madi,2021zhou-MDI,2018bunna}, free-space~\cite{2017Liao2,2020chenhuan,2021Avesani}, and underwater channels~\cite{2019jinxianmin}. Various quantum field networks have been reported worldwide~\cite{2010Chentengyun,2011Sasaki,2014wang,2019Dynes-network,2021Yang-network}. Interestingly, an integrated space-to-ground quantum network, based on a trusted-relay structure enabling multi-user secure communication over a total distance of 4600 km, was recently implemented~\cite{2021Chenyuao-network}. Even more recently, the record-breaking distances of QKD have been pushed to 511 km for field-deployed fiber~\cite{2021Chenjiupeng} and 605 km for fiber spool~\cite{2021Pittaluga} based on an efficient version of a measurement-device-independent QKD protocol~\cite{2012Lo} called twin-field QKD~\cite{2018Lucamarini,2018Maxiongfeng,2018wangxiangbin,2019Wangshuang}.   
	
	The security of QKD is provided by the principle of quantum physics, assuming that the features of real-life components conform to the theoretical models in the security proof~\cite{GLLP}. However, existing imperfections in practical implementations break these ideal assumptions, leaving several considerable vulnerabilities to eavesdropping by Eve. Indeed, multiple quantum hacking attacks~\cite{2018Qiang-hacking,2018Yoshino-hacking,2019Wei-hacking,2020Huanganqi-hacking,2020Pang-hacking} have been proposed by exploiting such realistic security loopholes (see~\cite{2020Xu} for a recent review on this topic).  
	
	In the current security proofs of QKD, a fundamental assumption is that the intensity of a quantum signal is not relate to its actual encoded state~\cite{GLLP}. The goal is to prevent Eve from learning the encoded bit by performing an unambiguous state discrimination attack~\cite{2013Tang}. Unfortunately, this key assumption cannot be guaranteed by state-of-the-art polarization-encoding modulators, which are mainly integrated using several fiber or silicon photonics components. This is because almost all of the above optical components, arising from physical structures, inevitably have some amount of polarization-dependent loss (PDL). For instance, according to Ref.~\cite{2017Sibson}, the PDL due to carrier-depletion modulators was approximately 1 dB.
	
	In this study, we experimentally characterized PDL in realistic polarization state preparation schemes and verified that PDL exists in fiber-~\cite{2019Agnesi,2019Li} and silicon-based polarization modulators (PMs)~\cite{2020Wei}. 
	Furthermore, we report a decoy-state BB84 QKD experiment that considers the PDL. Our demonstration exploits a novel theoretical proposal of Li et al.~\cite{2018Lichenyang}, which enables long-distance QKD through the post-selection of signals. We call this proposal a polarization-loss-tolerant  protocol. With the refined security proof, we successfully distributed secure key bits over different fiber links up to a 75 km. In contrast, no secure key bits can be generated using standard Gottesman-Lo-L{\"u}tkenhaus-Preskill (GLLP) analysis~\cite{GLLP}. The theoretical and experimental contributions are detailed below. 
	
	Theoretically, we combine the one-decoy-state method with the polarization-loss-tolerant protocol. This can significantly simplify the experimental complexity of the polarization-loss-tolerant protocol. Note that the one-decoy-state method has recently been proven to outperform the two-decoy-state method for almost all experimental settings, and only one decoy is easier to implement ~\cite{2018Rusca}. Thus, our analysis is crucial for implementing a polarization-loss-tolerant protocol and ensuring the security of practical QKD existing in the PDL. In addition, we quantify the security of QKD systems in the presence of PDL using a standard GLLP approach. This provides a quantitative observation for relating the security to specific values of PDL in PM. 
	
	Experimentally, we verified that PDL exists in recently proposed fiber- and silicon-based polarization modulation schemes. Furthermore, we performed the first decoy-state BB84 QKD demonstration using a homemade QKD system by considering the PDL. We quantified the PDL in state-preparation devices and considered it into the key rate formula. Using the polarization-loss-tolerant protocol, we successfully distributed secure key bits over up to 75 km of commercial fiber spool.
	
	The remainder of this paper is as follows. In Sec.~\ref{mismatch}, we present the one-decoy-state polarization-loss-tolerant protocol.  In Sec.~\ref{Experiment}, we describe our experimental setup and present the experimental results. Finally, we summarize our work in Sec.~\ref{conclusion}. 

	\section{Polarization-loss-tolerant protocol with one-decoy-state  method}\label{mismatch}
	\subsection{Original protocol}
	The key idea of the polarization-loss-tolerant protocol is that the photons unbalanced by the PDL can be randomly discarded. Hence, the final secret key is only extracted from the single-photon components whose density matrices are maximally mixed~\cite{2018Lichenyang}. In this manner, the destroyed assumption due to the PDL is restored. Furthermore, a post-selection scheme is introduced to reduce the consumption of error correction, and a higher secret key rate is obtained. With the refined security proof in the polarization-loss-tolerant protocol, the final secret key bits can be given as
	\begin{equation}
		\label{original-key-rate}
		R \geqslant q\left\{-\hat{Q}_{\mu} f\left(\hat{E}_{\mu}\right) H\left(\hat{E}_{\mu}\right)+\hat{Q}_{1}\left[1-H\left(e_{1}^{ph}\right)\right]\right\},	
	\end{equation}  
	where $q$ is the efficiency of the protocol,  $\hat{Q}_{1}$ and $e_{1}^{ph}$ are the gain and phase error rate of the single-photon states, respectively,  $\hat{Q}_{\mu}$ and $\hat{E}_{\mu}$ denote the gain and overall QBER of the signal states, respectively, $f(\hat{E}_{\mu})$ is the efficiency of error correction, and $H(x)$ is the binary Shannon entropy.
	
	The parameters required in Eq.~(\ref{original-key-rate}) can be estimated using the decoy-state technique~\cite{,2005wangxiangbin,2005Lo} for different polarizations, which can be summarized as  
	
	\begin{equation}
		\label{refined proof}
		\begin{aligned}
			\hat{Q}_{1} & = \min \left\{P \times\mu_{H} e^{-\mu_{H}}, \mu_{V} e^{-\mu_{V}}\right\} \times Y_{1}, \\
			Y_{1} & = \frac{Y_{1, H}+Y_{1, V}}{2}, \\
			e_{1}^{ph} & = \frac{Y_{1, D} e_{1, D}+Y_{1, A} e_{1, A}}{Y_{1, D}+Y_{1, A}}, \\
			\hat{Q}_{\mu} & = \frac{P \times Q_{\mu_H} + Q_{\mu_V}}{2} , \\
			\hat{Q}_{\mu} \hat{E}_{\mu} & = \frac{ P \times Q_{\mu_H} E_{\mu_H} + Q_{\mu_V} E_{\mu_V}}{2} ,
		\end{aligned}
	\end{equation}
	
	where $\mu_{M}$ with $M \in\{H,~V,~D,~A\}$ represent the intensities of the signal state prepared in a given polarization $M$. $P$ is the post-selection probability to compensate the single-photon components loss of the V base, given by $\mu_{V} e^{-\mu_{V}}=P \times \mu_{H} e^{-\mu_{H}}$.  $Y_{1, M}$ and $e_{1,M}$ denote the yield and QBER of the single-photon state prepared in the given polarization $M$, respectively. Moreover, $Q_{\mu_M}$ and $E_{\mu_M}$ are the gain and QBER of the signal states prepared in the given polarization $M$, respectively.

	\subsection{Parameter estimation using one-decoy-state method}
	In Ref.~\cite{2018Lichenyang}, the parameters required in Eq.~(\ref{original-key-rate}) were estimated using the two-decoy-state method, which has a relatively complex implementation. Here, we used the one-decoy-state method~\cite{2018Rusca} for parameter estimation, which significantly reduced the experimental complexity. Based on the framework presented in~\cite{2018Rusca}, the final secure key is given by  
	  \begin{equation}
		\begin{aligned}
			l\geqslant& s_{z, 0}^{L}+s_{z, 1}^{L}(1-h\left(e_{z,1}^{ph}\right)) \\
			&-leak_{EC}-6 \log _{2} \frac{19}{\varepsilon_{sec}}-\log _{2} \frac{2}{\varepsilon_{cor}},
		\end{aligned}
		\label{lfinite}
	\end{equation} \blk
	where $s_{z,0}^{L}$ is the lower bound of the detection counts by Bob given that Alice sent the vacuum pulses in the \red $z$ \blk basis, $s_{z, 1}^{L}$ is the analytical lower bound of the single-photon pulses in the \red $z$ \blk  basis, and $e_{z}^{ph}$ is the phase error rate. $ leak_{EC}$ is the number of announced bits in the error correction stage, and $\varepsilon_{sec}$ and $\varepsilon_{cor}$ are the secrecy and correctness criteria, respectively. 
	
	\begin{table*}[ht!]
		\centering
		\caption{Concrete descriptions for one-decoy-state polarization-loss-tolerant protocol.}
		\begin{tabular}{l}
			\hline \hline 
			$Definitions:$ \\
			$\lambda:$ basis choice, $ \lambda \in \{z,x\}$.\\
		    $k:$ intensity choice for signal and decoy state, $k \in \{\mu,\nu\} $.\\ 
			$M:$ polarization choice, $M \in \{H,V,D,A\}$.\\
			$k_M:$ intensity choice for given polarization $M$.\\
			$p_{\lambda}:$ probability choice for basis $\lambda$, $p_{\lambda} \in \{p_{z},(1-p_{z})\}$.\\
			$p_{k}:$ probability choice for intensity $k$, $p_{k} \in \{p_{\mu},p_{\nu}\}$.\\
			$P_{k}:$ post-selection probability choice for intensity $k$, $P_{k} \in \{P_{\mu},P_{\nu}\}$.\\
			$L_{\lambda}:$ polarization-dependent loss coefficient for basis $\lambda$.\\
			\quad \\
			$Measured\quad quantities:$\\
			$n_{z}:$ total number of detected pulses when Alice sends states in basis \red $z$ \blk  .\\
			$n_{\lambda,k_M}:$ number of detected pulses when Alice sends states in basis $\lambda$  and polarization $M$ with intensity $k$.\\
			$m_{\lambda,k_M}:$ number of error pulses when Alice sends states in basis $\lambda$  and polarization $M$ with intensity $k$.\\
			\quad \\
			$Statistical\quad fluctuations:$\\
			$\delta:$ statistics, $\delta\left(\chi, \varepsilon\right):=\sqrt{\chi \log \left(1 / \varepsilon\right) / 2}$.\\
			$n_{\lambda,k_M}^{\pm }:$ upper and lower bounds of $n_{\lambda,k_M}$, $n_{\lambda,k_M}^{\pm }=n_{\lambda, k_M}\pm \delta\left(n_{\lambda},\varepsilon_{1}\right)$.\\
			$m_{\lambda,k_M}^{\pm}:$ upper and lower bounds of $m_{\lambda,k_M}$, $m_{\lambda,k_M}^{\pm}=m_{\lambda, k_M} \pm \delta\left(m_{\lambda},\varepsilon_{2}\right)$.\\
			$\tau_{n,M}:$ $n$-photon-state probability for polarization $M$, $\tau_{n,M}=\sum_{k \in_{\mu,\nu}} p_{k} e^{-k_{M}} k_{M}^{n} / n !$.\\
			\quad\\
			$ Decoy-estimation\quad results:$\\
			$s_{z,0}^{L}:$ lower bound of vacuum events in basis \red $z$ \blk  according to Eq.~(\ref{Sz0L}).\\
			$s_{z,1}^{L}:$ lower bound of single-photon events in basis \red $z$ \blk  according to Eq.~(\ref{Sz1L}).\\
			$e_{z}^{ph}:$ phase error rate in the $Z$ basis according to Eq.~(\ref{pz1}).\\ 	
			\hline \hline
		\end{tabular}
		\label{parameter_estimation}
	\end{table*}    
	Due to the presence of the PDL, according to Eq.~(\ref{refined proof}), $\{s_{z,0}^{L},s_{z, 1}^{L},e_{z}^{ph}\}$ can be estimated from the measured quantities for different polarizations.
	In detail, let $s_{\lambda,n,M}$ be the number of detection counts measured by Bob given that Alice prepares $n$-photon states in basis $\lambda \in \{z,x\}$ and polarization $M$. In the asymptotic case, we obtained the number of detected pulses when Alice sends states in basis $\lambda$ and polarization $M$ with intensity $k\in\{\mu,\nu\}$ as 
	\begin{equation}
		n_{\lambda,k_M}^{*}=\sum_{n=0}^{\infty} p_{{k_M}|n} s_{\lambda,n,M},
	\end{equation} 
	where $p_{{k_M}|n}=\frac{p_{k}}{\tau_{n,M}} \frac{e^{-k_{M}} k_{M}^{n}}{n !}$ is the conditional probability of selecting intensity $k$ provided that Alice prepares an $n$-photon pulse in polarization $M$, and the subscript $*$ denotes the presence of an asymptotic case. Furthermore, $\tau_{n,M}=\sum_{k \in_{\mu,\nu}} p_{k} e^{-k_{M}} k_{M}^{n} / n ! $ is the probability that Alice prepares an $n$-photon pulse for polarization $M$, where $k_{M}$ represents the intensity of the state prepared in a particular polarization $M$. Correspondingly, let $v_{\lambda,n}$ be the number of errors detected by Bob when Alice sends an $n$-photon pulse and $m_{\lambda}=\sum_{n=0}^{\infty} v_{\lambda,n}$ is the total number of errors in the $\lambda$ basis. In the asymptotic case corresponding to different polarizations, the number of error pulses when Alice sends states in basis $\lambda$  and polarization $M$ with intensity $k$ can be obtained as
	\begin{equation}
		m_{\lambda,k_M}^{*}=\sum_{n=0}^{\infty} p_{{k_M}|n} v_{\lambda,n,M}.
	\end{equation} 
	
	Here, we adopt the observed counts in the   $z$ \blk  basis to distill the secret key. When Alice sends $n$-photon pulses in   $z$ \blk  basis, $n_{z}=\sum_{n=0}^{\infty} (s_{z,n,H}+s_{z,n,V}) $ is the all number of detection. Based on the polarization-loss-tolerant protocol, we have  
	\begin{equation}\label{Sz1}
		\begin{aligned}
			n_{z}&=\sum_{k \in_{\mu,\nu}} n_{z,k_H}\times P_{k} +n_{z,k_V}.
		\end{aligned}
	\end{equation} 
	Here, $P_k$ is the post-selection probability, which is helpful for obtaining a superior secure key length, particularly with a large PDL.   Since the detection events of $x$-basis are only used to estimate single-photon phase error rate, the post-selection probability on $x$ basis is needless. \blk The lower bound of the vacuum events $s_{z,0}^{L}$ can be achieved as follows:
	\begin{equation}\label{Sz0L}
		\begin{aligned}
			s_{z,0}^{L}&:= s_{z,0,H}^{L}+s_{z,0,V}^{L},
		\end{aligned}
	\end{equation}
	where $s_{z,0,H}^{L}(s_{z,0,V}^{L})$ is the lower bound of the vacuum events estimated by the set of detection events for polarization $H(V)$. Provided that the number of detection counts $n_{z,k_H}$ exceeds $n_{z,k_V}$, the lower bound of single-photon events $ s_{z, 1}^{L}$ can be found as follows:
	\begin{equation}\label{Sz1L}
		\begin{aligned}
			s_{z,1}^{L}:&=min[\tau_{1,H},\tau_{1,V}](\frac{s_{z,1,H}^{L}}{\tau_{1,H}}+\frac{s_{z,1,V}^{L}}{\tau_{1,V}}),
		\end{aligned}
	\end{equation}
	where $s_{z,1,H}^{L}(s_{z,1,V}^{L})$ is the lower bound of single-photon events estimated by the set of detection events for polarization $H(V)$, and $\tau_{1,H}= p_{\mu} \mu_{H} e^{-\mu_{H}}\times P_{\mu} +p_{\nu} \nu_{H} e^{-\nu_{H}}\times P_{\nu}$.  
	
	For the phase error rate, $\phi_{z,1}$ is estimated from the number of detections in the   $x$ \blk  basis~\cite{2014Lim}, which can be expressed as 
	  \begin{equation}\label{pz1}
		\phi_{z,1}^{ph} \leq \phi_{x,1}^{U}:=\frac{v_{x,1,D}^{U}+v_{x,1,A}^{U}}{s_{x,1,D}^{L}+s_{x,1,A}^{L}},
	\end{equation}  \blk                                        
	where $v_{x,1,D}^{U}(v_{x,1,A}^{U})$ is the upper bound of the single-photon error events by the set of error detection events for polarization $D(A)$.  The concrete descriptions and formulas are summarized in Table~\ref{parameter_estimation}. More details on the one-decoy-state polarization-loss-tolerant protocol can be found in  Appendix \ref{AppendixB}.       
	
	\section{Experiment and discussion}\label{Experiment}
	
	\subsection{Setup}\label{setup}
\begin{figure*}
		\centering
		\includegraphics[width=0.9\textwidth]{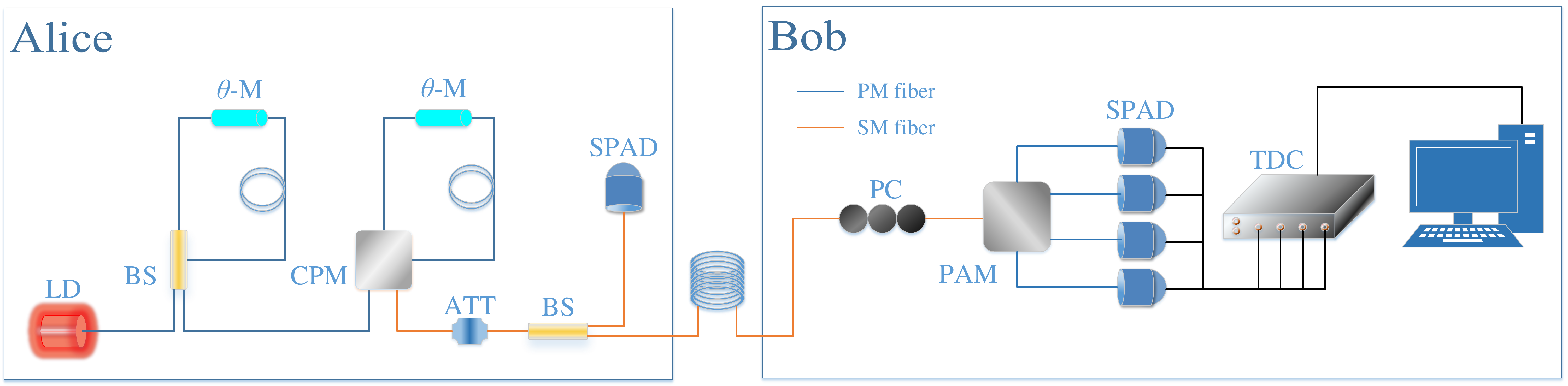}
		\caption{(Color online) Schematic diagram of BB84 QKD experimental setup. LD: 1550 nm commercial laser source; BS: beam splitter; $\theta$-M: phase modulator; CPM: customized polarization module; ATT: variable attenuator; PM fiber, polarization-maintaining fiber; SM fiber, single mode fiber; PC: polarization controller; PBS: polarized beam splitter; SPAD: single-photon avalanche detector; TDC: time-to-digital converter.  
		}\label{fig1}
	\end{figure*}

	We implemented the polarization-loss-tolerant protocol using a homemade polarization-encoding QKD system~\cite{2021Madi}. A schematic diagram of our setup is shown in Fig.~\ref{fig1}. Alice generated laser pulses at a clock frequency rate of 50 MHz using a commercial laser source (LD, WT-LD, Qasky Co. LTD). The pulses were coupled into a Sagnac-based intensity modulator actively modulating the intensities of each pulse for the decoy-state method.  Subsequently, the laser pulses entered a Sagnac-based polarization modulator (Sagnac-PM)~\cite{2019Li}, which  modulates four polarization states for the BB84 protocol. Then, the encoded pulses are were attenuated by a variable optical attenuator (VOA) to single-photon levels.
	
	The receiver Bob possessed a PC to actively compensate for the deflection of polarization during transmission over fiber spools.   The QBER of the system was used as the error signal for the active compensation. \blk The received pulses were de-encoded using a customized polarization analysis module (PAM) integrated with a 90/10 beam splitter and two polarization-maintaining polarized beam splitters. The photons were detected using four InGaAs single-photon avalanche detectors (SPADs, WT-SPD2000, Qasky Co. LTD) with a detection efficiency of $8.8\%$, dark count rate of $10^{-6}$ per pulse, and an after pulsing probabilities of 3\%. The detection events were recorded using a time-to-digital converter (TDC, quTAG100, GmbH). An optical misalignment error of approximately $1\%$ was achieved by carefully calibrating the system.

	\subsection{Quantifying PDL}
	We quantified the PDL in the source by measuring the intensity of each polarization generated by the Sagnac-PM. The measurement process was as follows. We first calibrated the expected voltages for different polarizations and determined that the voltages $\{0,~V_\pi,~0.5V_\pi, ~-0.5V_\pi\}$ modulate the expected polarization $\{H,V,D,A\}$, where $V_\pi=3.8$ V. Our calibration follows a custom procedure where we scan the applied voltages of a phase modulator and record the photon detection counts D$_1$ and D$_2$. Then V$_\pi$ is determined when the maximal visibility of $V=(D_1-D_2)/(D_1+D_2) $ is reached. Subsequently, the Sagnac-PM was directly connected to a high-precision optical power meter. Alice scanned the voltages applied to her Sagnac-PM and recorded the mean power of the optical power meter. These values were denoted by $P_M$. The polarization-dependent loss $L_{z(x)}$ for basis $z(x)$ was then calculated as follows:
	
	\begin{equation} 
		L_{z(x)}= P_{H(D)}-P_{V(A)}.
	\end{equation}
	
	 For comparison, we also measured the PDL in recently proposed PM schemes, including an all-fiber self-compensating polarization encoder (AS-PM)~\cite{2019Agnesi} and a silicon-based PM (Silicon-PM)~\cite{2020Wei}. The AS-PM was re-engineered with commercially available products, including a circulator and polarized beam splitter (Optizone Ltd.), phase modulator (iXblue Ltd.), and polarization controller (Thorlabs, Inc.). The Silicon-PM was manufactured by the standard fabrication service offered by IMEC foundry. The measurement process was similar to that for the Sagnac-PM.   When we measured the PDL of AS-PM and Silicon-PM, the launch power of laser pulse is set to -26.557 dBm. Since the loss of Sagnac-PM is larger than previous schemes (this rises from our customized CPM), we enhanced the laser power to -20.408 dBm for making the responsivity of power meter in the linear region. \blk All measured power and corresponding $L$ values are listed in Table~\ref{table1}. The table shows that all realistic PMs exhibited a PDL. In particular, the PDL was as large as 2.24 dB for a Silicon-based PM.   In the table, we   listed the results of the PDL in $x$ basis. These can be applied in other protocol~\cite{2014Lim} where the $x$ basis is used to  generate key bits.  We also noticed that the PDL of Sagnac-PM is larger than that of a fiber-based polarization modulation. This is arsing from the imperfections of our in-house-designed, customized polarization module (CPM) in the Sagnac-PM.  A detailed analysis can be found in Appendix~\ref{Sagnac-PM PDL}. \blk

\begin{table}[ht!]
		\centering
		\caption{Power and PDL for different polarization encoding modules.   $IL$ (dB) denotes an average insertion loss. \blk The output power $P_{M}$(dBm) represents the mean power for polarization $M$. $L_{z(x)}$ is the restored polarization loss in the $z(x)$ basis.   }
			\scalebox{0.85}{\begin{tabular}{lccccccc}
			\hline \hline
			Module&$IL$&$P_H$&$P_V$&$P_D$&$P_A$&$L_z$&$L_x$\\
			\hline Sagnac-PM&23.4&$-43.488$&$-44.324$&$-43.907$&$-43.832$&$0.836$&$-0.075$\\
			AS-PM&8.2&$-34.775$&$-34.658$ &$-34.666$&$-34.853$&$-0.117$&$0.187$\\
			Silicon-PM&5.3&$-31.08$&$-33.32$&$-32.02$&$-30.82$&$2.24$&$-1.36$\\
			\hline \hline	
		\end{tabular}}
		\label{table1}
	\end{table} 
	
	\subsection{Implementation of polarization-loss-tolerant protocol}

	We implemented the polarization-loss-tolerant protocol over commercial fiber lengths of $25$ km, $50$ km, and $75$ km. For each distance, we optimize the implementation parameters through a numerical simulation tool, including the intensities of the signal and decoy states, the probabilities of sending them, and the post-selection probability $P_k$. The optimization routine was similar to that in Ref.~\cite{,2018Lichenyang}, except we used the one-decoy-state method.
	
\begin{figure}
		\centering
		\includegraphics[width=0.45\textwidth]{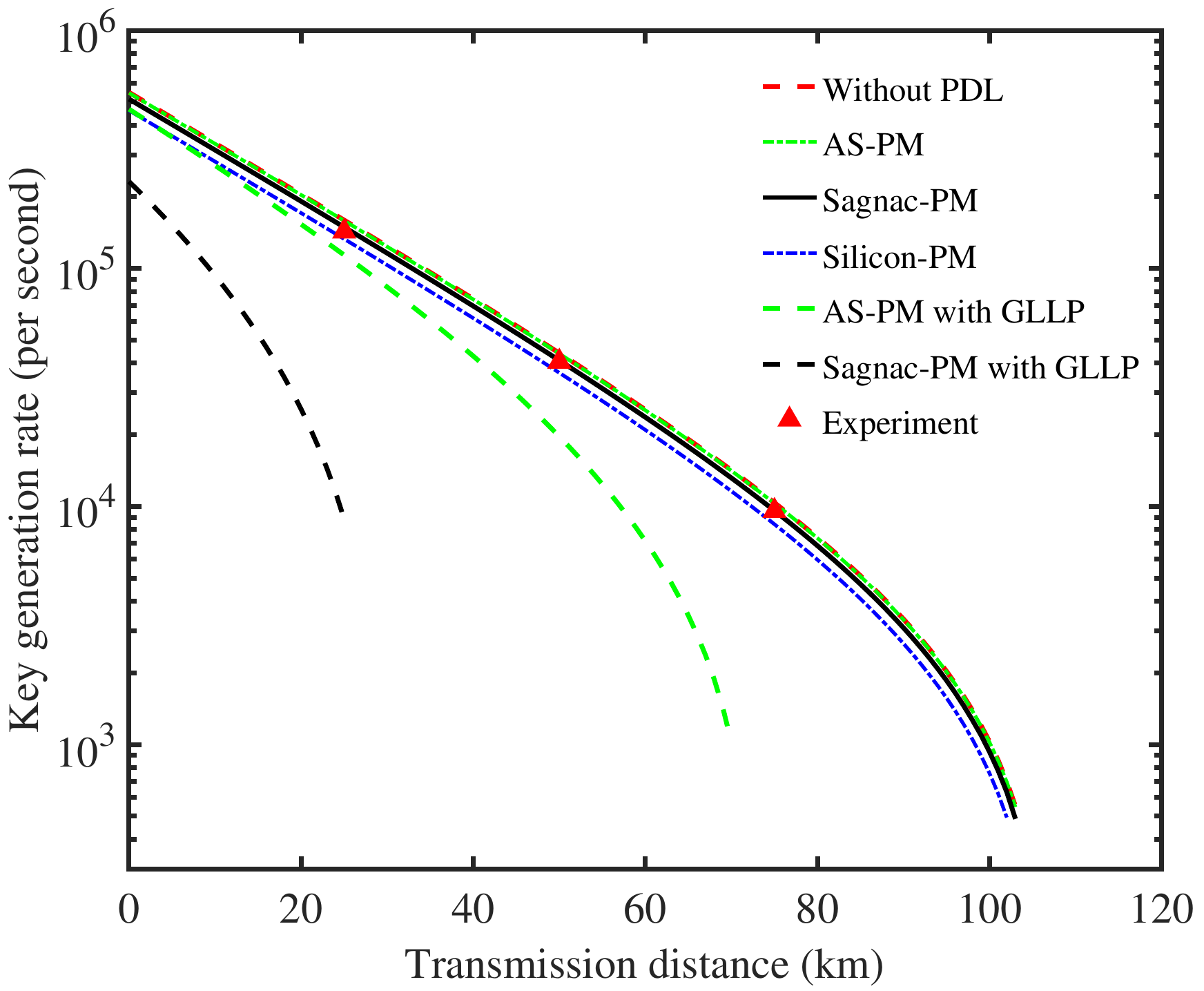}
		\caption{ Secure key rate with PDL in a practical setting. The black and green and blue dotted curves represent the key rates using the one-decoy-state polarization-loss-tolerant protocol. The green and black dashed curves denote the one-decoy-state BB84 QKD with the standard GLLP analysis for PDL. The carmine triangles represent the obtained experimental results.    
		}\label{fig2}
	\end{figure}

	For each distance, we sent a total number of $N=10^{10}$ pulses. As indicated in Table~\ref{parameter_estimation}, we collected the counts for different polarizations, and the details are provided in Appendix  ~\ref{experimental-results}\blk. By inputting the experimental counts into the one-decoy-state method presented in Sec.~\ref{mismatch}, we obtained the experimental results listed in Table~\ref{table2} and plotted in Fig.~\ref{fig2}.   In Table~\ref{table2}, the obtained QBERs for each distance are monotonically increasing because of calibration systematic error, which ranges from $0.8$ to $1.1\%$. \blk With the polarization-loss-tolerant protocol, we achieved a secure key rate of 9.58 kbps at a distance of up to 75 km. The security of these keys considers the PDL in the PM.
	
	\begin{table*}[ht!]
		\centering
		\caption{Implementation parameters and experimental results. L and Loss are the channel lengths and channel loss, respectively. $N$ is the total number of sent pulses. $\mu(\nu)$ is the intensity of signal(decoy) state, $p_{\mu}(p_{\nu})$ is the send probability of the signal(decoy) state, and $P_{\mu}(P_{\nu})$ is the post-selection probability for the signal(decoy) state. \blk $ e_{z}^{ph}$ is estimated phase error rate. $\hat{E}_{\mu}$ is obtained QBER. $l$ denotes the final key rate.}
		\begin{tabular}{cccccccccccc}
			\hline
			\multicolumn{2}{c}{Channel} & \multicolumn{7}{c}{Parameter} & \multicolumn{3}{c}{Results}\\ \hline
			L(km)& Loss(dB)&$N$&$\mu$&$\nu$&$p_{\mu}$&$p_{\nu}$&$P_{\mu}$&$P_{\nu}$&$e_{z}^{ph}$&$\hat{E}_{\mu}$&$l$\\ 
			\hline
			25&4.720&$1.0020\times10^{10}$&$0.626$&$0.157$&$0.85$&$0.15$&$0.9196$&$0.8476$&$2.90\%$&$1.04\%$&$1.43\times10^5$\\ 
			\hline
			50&9.812&$1.0026\times10^{10}$&$0.619$&$0.155$&$0.78$&$0.22$&$0.9189$&$0.8475$&$5.97\%$&$1.14\%$&$4.06\times10^4$\\ 
			\hline
			75&14.970&$1.0001\times10^{10}$&$0.611$&$0.153$&$0.67$&$0.33$&$0.9171$&$0.8470$&$8.63\%$&$1.11\%$&$9.58\times10^3$\\ 
			\hline
		\end{tabular}
		\label{table2}
	\end{table*} 
	
	To illustrate the implications of our results, as shown in Fig.~\ref{fig2}, we also plotted the simulation results following a standard GLLP analysis with PDL~\cite{GLLP}. That is, we considered the PDL as small source basis-dependent flaws and applied it to the standard GLLP key rate formula, as in the study on GLLP analysis for state modulation flaws~\cite{,2014Tamaki,2015Xu,2016Tangzhiyuan}. A detailed analysis can be found in Appendix \ref{Appendix-GLLP}. The simulation exploited experimental parameters obtained in our setup and the PDL listed in Table~\ref{table1}. Fig. 2 shows that with increasing $L_z$, the key generation rate rapidly decreased using a standard GLLP analysis. In particular, the key generation rate dropped to zero with $L_z=2.24$ dB, obtained in the Silicon-PM. The maximal tolerant distance was 25 km for our setup ($L_z=0.836$ dB) using a previous standard GLLP analysis. In contrast, our security analysis ensures that the QKD setup is secure over 100 km, implying that for the 75-km demonstration, not even a single bit could be extracted using the previous GLLP analysis.

	 In our experiment, we also measure  the PDL of SPADs, which is equal to zero in common sense. A value of less than 0.14 dB is obtained. However, since the PDL of SPADs results in a polarization dependency on the detection efficiency of detectors, it can be treated as a kind of detector efficiency flaws. Hence it did not influence the experimental demonstration of the polarization-loss-tolerant protocol, which focuses on the source flaws. In fact, in our previous work~\cite{2019Wei-hacking}, we have analyzed the impact of the polarization-dependent efficiencies on superconducting nanowire single-photon detector, and proposed some solutions to remove such a loophole.
	\blk
	
	\section{Conclusion}\label{conclusion}
	
	In summary, we demonstrated a decoy-state BB84 QKD experiment considering PDL. Following the one-decoy-state polarization-loss-tolerant protocol, we successfully generated secure key bits over different fiber links of up to 75 km. In contrast to previous experiments, which ignored the PDL in POL, the proposed study showed the feasibility of distributing secure key bits in the presence of PDL. Although we demonstrated the polarization-loss-tolerant protocol using a homemade polarization-encoding system, this method could be easily applied to other BB84-QKD systems~\cite{2020Agnesi}. Furthermore, it will be interesting to combine our results with other types of QKD systems, such as measurement-device-independent or twin-field QKD systems.
	
	\section{Acknowledgments}
	We thank F. Xu and W. Li for providing the silicon chip. This study was supported by the National Natural Science Foundation of China (No. 62171144 and No. 62031024) and the Guangxi Science Foundation (Grant No. 2017GXNSFBA198231).
	
	\appendix

	\section{Parameter estimation using one-decoy-state method}\label{AppendixB}
	In this section, we present our one-decoy-state parameter estimation for a polarization-loss-tolerant protocol. The finite-data size was also included using the framework in Ref.~\cite{2014Lim}.
	
	The total number of detections in the $\lambda$ basis is given by $n_{\lambda}=\sum_{n=0}^{\infty} s_{\lambda, n}$ ($ \lambda \in {z,x}$), where $s_{\lambda,n}$ are the detection events when Alice sends an $n$-photon pulse. When PDL is present, the protocol assigns the detection counts corresponding to each polarization state $n_{\lambda,M}$ separated from the data set $n_{\lambda}$, where $M\in\{H, V, D, A\}$. In the asymptotic limit, the number of detections with a specific intensity $k\in\{\mu,\nu\}$ is given by
	\begin{equation}
		n_{z,k_M}^{*}=\sum_{n=0}^{\infty} p_{{k_M}|n} s_{z,n,M}.
	\end{equation} 
	Here, $p_{{k_M}|n}$ is the conditional probability, which can be expressed as $p_{{k_M}|n}=\frac{p_{k}}{\tau_{n,M}} \frac{e^{-k_{M}} k_{M}^{n}}{n !}$, where $\tau_{n,M}=\sum_{k \in_{\mu,\nu}} p_{k} e^{-k_{M}} k_{M}^{n} / n ! $ is the probability that Alice prepares an $n$-photon pulse for polarization $M$. Here, $p_{k}$ is the probability of choosing the signal or decoy state, and $k_{M}$ represents the intensities $k$ of the state prepared in a given polarization $M$. 
	
	However, the observed data $n_{\lambda,k_M}$ are different from the corresponding asymptotic case when considering a finite-statistic scenario. By employing Hoeffding’s inequality~\cite{1963Hoeffding} for independent variables to bind the fluctuation, the experimental data satisfy
	\begin{equation}
		\left|n_{\lambda,k_M}^{*}-n_{\lambda,k_M}\right| \leq \delta\left(n_{\lambda}, \varepsilon_{1}\right),
	\end{equation}
	with the probability at least $1-2 \varepsilon_{1}$, where $\delta\left(n_{\lambda}, \varepsilon_{1}\right):=\sqrt{n_{\lambda} \log \left(1 / \varepsilon_{1}\right) / 2}$. The above equation allows us to obtain the upper and lower bounds of the counts $n_{\lambda,k_M}^{*}$ as follows:
	\begin{equation}
		\begin{aligned}
			&n_{\lambda,k_M}^{*} \leqslant n_{\lambda,k_M}+\delta\left(n_{\lambda}, \varepsilon_{1}\right)=n_{\lambda,k_M}^{+} \\
			&n_{\lambda,k_M}^{*} \geqslant n_{\lambda,k_M}-\delta\left(n_{\lambda}, \varepsilon_{1}\right)=n_{\lambda,k_M}^{-}.
		\end{aligned}
	\end{equation}
	
	For the error detection events, we consider that the value $v_{\lambda,n}$ is the number of errors detected by Bob when Alice sends an $n$-photon pulse and $m_{\lambda}=\sum_{n=0}^{\infty} v_{\lambda,n}$ is the total number of errors in the $\lambda$ basis. In the asymptotic case corresponding to the polarization, we have 
	\begin{equation}
		m_{\lambda,k_M}^{*}=\sum_{n=0}^{\infty} p_{{k_M}|n} v_{\lambda,n_M}, \forall k \in\{\mu,\nu\}.
	\end{equation}
	In reference to the previous case, we can determine the difference between the experimental values $m_{\lambda,k_M}$ and the corresponding asymptotic case $m_{\lambda,k_M}^{*}$, as follows:
	\begin{equation}
		\left|m_{\lambda,k_M}^{*}-m_{\lambda,k_M}\right| \leq \delta\left(m_{\lambda}, \varepsilon_{2}\right)
	\end{equation}
	with the probability of at least $1-2\varepsilon_{2}$.
	
	Based on an estimation method proposed in~\cite{2018Rusca,2014Lim}, the lower bound of the vacuum counts in the $\lambda$ basis can be expressed as follows:
	\begin{equation}\label{S0}
		\begin{aligned}
			s_{\lambda,0,M} \geq s_{\lambda, 0,M}^{L}&:=\frac{\tau_{0,M}}{\mu_{M}-\nu_{M}}\left(\mu_{M} n_{\lambda,\nu_M}^{-}-\nu_{M} n_{\lambda,\mu_M}^{+}\right),
		\end{aligned}
	\end{equation}
	and the lower bound of the single-photon counts for polarization $M$ on the basis of $\lambda$ is given by 
	\begin{equation}\label{S1}
		\begin{aligned}
			s_{\lambda,1,M} \geq s_{\lambda,1,M}^{L}:&=\frac{\tau_{1,M}\mu_{M}}{\nu_{M}\left(\mu_{M}-\nu_{M}\right)}(n_{\lambda,\nu_M}^{-}\\
			&-\frac{\nu_{M}^{2}}{\mu_{M}^{2}} n_{\lambda,\mu_M}^{+}-\frac{\left(\mu_{M}^{2}-\nu_{M}^{2}\right)}{\mu_{M}^{2}} \frac{s_{\lambda, 0,M}^{U}}{\tau_{0,M}}),
		\end{aligned}
	\end{equation}
	where $s_{\lambda,0,M}^{U}$ is the upper bound of the vacuum counts through the error events that can be bound by $s_{\lambda,0,M}^{U}:=2 \tau_{0,M} \frac{e^{k_{M}}}{p_{k}} m_{\lambda,k_M}+ \delta\left(n_{\lambda}, \varepsilon_{1}\right)$.
	
	Considering a specific scenario, the following formula can be used to estimate the phase error in the   $z$ \blk  basis ~\cite{2010Fung}:
	\begin{equation}\label{pz}
		e_{z,1}^{ph} \leq e_{x,1}^{U}:=\frac{v_{x, 1}^{U}}{s_{x, 1}^{L}}+\gamma\left(\varepsilon_{s e c}, \frac{v_{x, 1}^{U}}{s_{x, 1}^{L}}, s_{z, 1}, s_{x, 1}\right)
	\end{equation}
	where
	\begin{equation}
		\begin{aligned}
			\gamma(a, b, c, d) \\
			=&\sqrt{\frac{(c+d)(1-b) b}{c d \log 2} \log _{2}\left(\frac{c+d}{c d(1-b) b} \frac{21^{2}}{a^{2}}\right)}.
		\end{aligned}
	\end{equation} 
	By applying the result in~\cite{2014Lim}, the upper bound of the number of single-photon error events in the   $x$ \blk  basis for polarization $D$ is given by 
	\begin{equation}\label{vx1}
		\begin{aligned}
			v_{x, 1,D}\leq v_{x, 1,D}^{U}&=\frac{\tau_{1,D}}{\mu_{D}-\nu_{D}}\left(m_{x, \mu_D}^{+}-m_{x, \nu_D}^{-}\right).
		\end{aligned}
	\end{equation} 
	Similarly, the upper bound of the single-photon error events $v_{x,1,A}^{U}$ can be obtained. Combining with Eq.~(\ref{S1}), we obtain:
	\begin{equation}
		\begin{aligned}
			v_{x,1}^{U}&= v_{x,1,D}^{U}+v_{x,1,A}^{U},\\ s_{x,1}^{L}&= s_{x,1,D}^{L}+s_{x,1,A}^{L}.
		\end{aligned}
	\end{equation} 
	  \section{Concise analysis of the source of Sagnac-PM PDL}
\label{Sagnac-PM PDL}
The Sagnac-PM has a larger PDL than that of a fiber-based PM to conform our customized polarization module (CPM) in the Sagnac-PM, as shown in Fig.~\ref{fig1}. We experimentally quantified the parameters including the splitter ratio $\alpha=0.872$ and the orthogonal deviation angle  $\theta=0.091$ and found that these specific parameters of the CPM are larger than that of a standard commercial component. This would be the main reason for the PDL of Sagnac-PM being considerably larger than that in a fiber-based system. The detailed analysis is as follows: 

When there is a deviation angle  $\theta$ between orthogonal components $|H\rangle$ and $|V\rangle$, the output light from The CPM can be expressed as 
\begin{equation}
	\left|E_{1}\right\rangle=A_{1} e^{i \omega}|H\rangle,
\end{equation}
and
\begin{equation}
	\left|E_{2}\right\rangle=\alpha A_{1} e^{i \omega+\varphi} \sin \theta|H\rangle+\alpha A_{1} e^{i \omega+\varphi} \cos \theta|V\rangle,
\end{equation}
where $\varphi$ is the encoded phase. Finally, the mean intensity of the light can be expressed as
\begin{equation}
	I=\int_{0}^{2 \pi}\left(\left\langle E_{1}|+\left\langle E_{2}|\right)\left(|E_{1}\right\rangle+|E_{2}\right\rangle\right) d \omega.
\end{equation} 
Then, the PDL of Sagnac-PM in z basis is given by
\begin{equation}
	\mathrm{L}_{z}=10 \log \frac{I(H)}{I(V)}.
\end{equation}
Using the measured data, we get the theoretical value $L_{z}=0.780$ dB, which is close to the measured value of PDL of Sagnac-PM ($L_{z}=0.836$ dB).

\blk	
	\section{Detailed experimental results}
	\label{experimental-results}
	Table~\ref{experimental_data} details the experimental results.
	
	\begin{table*}[ht!]
		\centering
		\caption{Experimental raw counts.}
		\label{experimental_data}
		\begin{tabular}{ccccccccc}
			\hline \hline Distance& $n_{x, \mu}$ & $n_{x, \nu}$ & $n_{z, \mu}$ & $n_{z, \nu}$ & $m_{x, \mu}$ & $m_{x, \nu}$ & $m_{z, \mu}$ & $m_{z, \nu}$ \\
			\hline $25 \mathrm{~km}$ & 1302170 & 65577 & 87895811 & 3945456 & 13094 & 945 & 889766 & 61921 \\
			\hline $50 \mathrm{~km}$ & 377232 & 26907 & 31562239 & 2278091 & 4933 & 690 & 346762 & 38516 \\
			\hline $75 \mathrm{~km}$ & 87033 & 13457 & 8632164 & 1074008 & 2137 & 858 & 84805 & 20998 \\
			\hline  \\
			\hline \hline ~ & $n_{x, \mu_D}$ & $n_{x, \mu_A}$ & $n_{x, \nu_D}$ & $n_{x, \nu_A}$ & $n_{z, \mu_H}$ & $n_{z, \mu_V}$ & $n_{z, \nu_H}$ & $n_{z, \nu_V}$ \\
			\hline $25 \mathrm{~km}$ & 610001 & 692169 & 37460 & 28117 & 45962402 & 41933409 & 2485621 & 1459835  \\
			\hline $50 \mathrm{~km}$ & 192261 & 184971 & 15120 & 11787 & 16950343 & 14611896 & 1441686 & 836405  \\
			\hline $75 \mathrm{~km}$ & 41745 & 45288 & 7455 & 6002 & 4633555 & 3998609 & 635057 & 438951 \\
			\hline  \\
			\hline \hline ~ & $m_{x, \mu_D}$ & $m_{x, \mu_A}$ & $m_{x, \nu_D}$ & $m_{x, \nu_A}$ & $m_{z, \mu_H}$ & $m_{z, \mu_V}$ & $m_{z, \nu_H}$ & $m_{z, \nu_V}$ \\
			\hline $25 \mathrm{~km}$ & 8591 & 4503 & 715 & 230 & 285548 & 604218 & 19337 & 42584 \\
			\hline $50 \mathrm{~km}$ & 1732 & 3201 & 208 & 482 & 160264 & 186498 & 16339 & 22177 \\
			\hline $75 \mathrm{~km}$ & 1561 & 576 & 661 & 197 & 37191 & 47614 & 7422 & 13576 \\
			\hline 
		\end{tabular}
	\end{table*}

	\section{Security bounds against PDL using standard GLLP analysis}
	\label{Appendix-GLLP}
	
	In this section, we discuss how we can bound information leakage caused by PDL using standard GLLP security analysis. We consider the PDL in state-preparation devices as a type of source flaw. Hence, the key rate formula is similar to that in Eq.~(\ref{original-key-rate}) in the main text, except that the phase error rate needs to include the correction due to source flaws. Based on the GLLP analysis, PDL can be quantified using the so-called quantum coin $\Delta$, which is given by
	\begin{equation}\label{kuang}
		\Delta=\frac{1-F\left(\rho_{z}, \rho_{x}\right)}{2},
	\end{equation} 
	where $F\left(\rho_{z}, \rho_{x}\right)$ is the fidelity of the density matrices for the   $z$ \blk  and   $x$ \blk  bases. The balance of a quantum coin quantifies the basis-dependent flaws of Alice's single-photon components or the ability to discriminate the bases that Eve possesses. For simplicity, we introduce the idea of an entanglement-based scenario to provide an imperfect parameter value, which is equivalent to a prepare-and-measure protocol. Here, Alice first generates an entangled state as follows:
	\begin{equation}\label{ABz}
		\left|\Upsilon_{z}\right\rangle_{A B}=\sqrt{\frac{1}{l_{z}+1}}\left|0_{z}\right\rangle_{A} \otimes\left|0_{z}\right\rangle_{B}+\sqrt{\frac{l_{z}}{l_{z}+1}}\left|1_{z}\right\rangle_{A} \otimes\left|1_{z}\right\rangle_{B}
	\end{equation} 
	and sends System B to Bob. Here, the coefficient $l_z$ depends on the polarization-dependent loss $L_z$, which can be expressed by $ l_z = 10^{-L_{z}/10}$. In the virtual protocol, Alice can measure system A after Bob detects and Eve makes a disturbance. In Eq.~(\ref{ABz}), we consider the PDL, from which the coefficient of the state is related to $L_{z}$, satisfying normalization. Similarly, for each   $x$ \blk  basis emission, Alice prepares the entangled states as follows:
	\begin{equation}\label{ABx}
		\left|\Upsilon_{x}\right\rangle_{A B}=\sqrt{\frac{1}{l_{x}+1}}\left|0_{x}\right\rangle_{A} \otimes\left|0_{x}\right\rangle_{B}+\sqrt{\frac{l_{x}}{l_{x}+1}}\left|1_{x}\right\rangle_{A} \otimes\left|1_{x}\right\rangle_{B}
	\end{equation} 
	and sends System B to Bob. The coefficient $l_{x}$ depends on the PDL on the   $x$ \blk  basis. Evidently, the states $\left|\Upsilon_{z}\right\rangle_{A B}$ and $\left|\Upsilon_{x}\right\rangle_{A B}$ are no longer equal because of the imperfect state preparation. Furthermore, by introducing a quantum coin, Alice prepares an entangled state following ~\cite{2009Koashi}  
	\begin{equation}
		\begin{aligned}
			|\Gamma\rangle_{C A B} &=\frac{1}{2}\left[| 0_{x} \rangle_{ C } \left(\left|\Upsilon_{z}\right\rangle_{A B}\right.\right.\\
			&+\left|\Upsilon_{x}\right\rangle_{A B})+\left|1_{x}\right\rangle_{C}\left(\left|\Upsilon_{z}\right\rangle_{A B}-\left|\Upsilon_{x}\right\rangle_{A B}\right)],
		\end{aligned}
	\end{equation} 
	where system C is "a quantum coin,” determining that each signal is encoded on a   $z$ \blk  or   $x$ \blk  basis. If the quantum-coin system collapses into the state $\left|1_{x}\right\rangle_{C}$, we can obtain the probability quantifying how well the basis dependence of Alice's and Bob's single-photon pairs, so that we have
	\begin{equation}
		\begin{aligned}
			\Delta  =\operatorname{Prob}\left(X_{C}=-1\right)&=\left.\left.\right|_{C}\left\langle 1_{x} \mid \Gamma\right\rangle_{C A B}\right|^{2} \\
			&=\frac{1}{2}(1-\frac{(1+\sqrt{l_{z}})(1+\sqrt{l_{x}})}{\sqrt{(l_{z}+1)(l_{x}+1)}}).
		\end{aligned}
	\end{equation}
	
	In our QKD system, with $L_{z(x)}=0.836(-0.075)$, we have $\Delta=5.82\times10^{-4} $. Thus, based on Eq.~(\ref{kuang}), the fidelity $F\left(\rho_{z}, \rho_{x}\right)=1-1.16\times10^{-3}$. In the GLLP analysis, the basis-dependent flaws of Alice’s signals associated with single-photon events can be enhanced in principle by Eve by exploiting the channel loss; thus, $\Delta$ is replaced by $\Delta^{\prime}$ as follows: 
	\begin{equation}
		\Delta^{\prime}=\frac{\Delta}{Y_{1}},
	\end{equation}
	where $Y_{1}$ is the yield of the $1$-photon pulses. The revised phase error rate can be expressed as 
	\begin{equation}
		\bar{e}_{z}^{ph} \leqslant e_{x,1}^{U}+4 \Delta^{\prime}+4 \sqrt{\Delta^{\prime} e_{x,1}^{U}}+\epsilon_{ph}.
		\label{phase-error}
	\end{equation}
	By substituting Eq.~(\ref{phase-error}) into Eq.~(\ref{original-key-rate}), we obtain the final key rate using the standard GLLP approach while considering the PDL. The simulation results are presented in Fig.~\ref{fig2}.

	
%

\end{document}